\documentclass[11pt, letterpaper, logo, onecolumn, copyright, numbering]{minimax}
\usepackage{etoolbox}

\usepackage[authoryear, sort&compress, round]{natbib}

\usepackage[inkscapeformat=png]{svg}

\usepackage[most, breakable, skins]{tcolorbox}
\usepackage{academicons}

\tcbuselibrary{skins}
\usepackage{lipsum}
\usepackage{tabularx}
\usepackage{afterpage}
\usepackage{booktabs}
\usepackage{url}
\usepackage{subcaption}
\usepackage{makecell}
\usepackage{multirow}
\usepackage{multicol} 
\usepackage{array}
\usepackage{float}
\usepackage{listings, listings-rust}
\usepackage{fontawesome5}
\usepackage{amssymb,graphicx}
\usepackage[dvipsnames]{xcolor}
\usepackage{hyperref}
\usepackage{cleveref}
\usepackage{longtable}
\usepackage{graphicx}
\usepackage{pdflscape}
\usepackage{adjustbox}
\usepackage{tikz}
\usetikzlibrary{calc,positioning,chains,shapes,arrows,fit,decorations.pathmorphing,patterns,fadings,shadows,patterns.meta,arrows.meta}
\usepackage{wrapfig}
\usepackage{dialogue}
\usepackage{algorithm}
\usepackage{algorithmic}
\usepackage{colortbl}
\usepackage{mdframed}
\usepackage{caption}

\usepackage{listings} 
\usepackage{CJKutf8}
\usepackage{tcolorbox} 
\usepackage[dvipsnames]{xcolor}
\usepackage{multicol}    

\usepackage{amsmath,amsfonts,bm}

\def\eqref#1{equation~\ref{#1}}

\def\1{\bm{1}}

\DeclareMathAlphabet{\mathsfit}{\encodingdefault}{\sfdefault}{m}{sl}
\SetMathAlphabet{\mathsfit}{bold}{\encodingdefault}{\sfdefault}{bx}{n}

\definecolor{ababcol}{HTML}{F14738}
\definecolor{myhailuo2}{HTML}{F97669}
\definecolor{querycol}{HTML}{7964E8}
\definecolor{goldanswercol}{HTML}{FFB43B}
\definecolor{otherscol}{HTML}{FC5BCF}
\definecolor{myhailuo3light}{HTML}{FFA9FA}
\definecolor{myhailuo1dark}{HTML}{FC8900}
\definecolor{myhailuo2dark}{HTML}{F14738}
\definecolor{myhailuo3dark}{HTML}{D12AAA}
\definecolor{myhailuo4dark}{HTML}{4C4DC2}

\definecolor{myhailuo1}{HTML}{FFB43B}
\definecolor{myhailuo2}{HTML}{F97669}
\definecolor{myhailuo3}{HTML}{FC5BCF}
\definecolor{myhailuo4}{HTML}{7964E8}

\definecolor{myhailuo1light}{HTML}{FFD085}
\definecolor{myhailuo2light}{HTML}{FFA19F}
\definecolor{myhailuo3light}{HTML}{FFA9FA}
\definecolor{myhailuo4light}{HTML}{BDACFB}

\colorlet{myorange}{Orange!20}
\colorlet{mygreen}{LimeGreen!25}
\colorlet{myyellow}{Yellow!30}
\colorlet{myblue}{CornflowerBlue!25}
\colorlet{mybrown}{RawSienna!25}
\colorlet{mypurple}{Orchid!25}
\colorlet{myred}{Red!60}
\colorlet{myorangefull}{YellowOrange!60}
\colorlet{mybrownfull}{RawSienna!60}

\colorlet{myorangethick}{Orange!40}
\colorlet{mygreenthick}{LimeGreen!50}
\colorlet{myyellowthick}{Yellow!60}
\colorlet{mybluethick}{CornflowerBlue!50}

\tcbset{
    showcase/.style={
        fonttitle=\large,
        colback=white!20,  
        colframe=black,   
        coltitle=white,   
        boxrule=0.5mm,    
        arc=2mm,          
        outer arc=2mm,    
        left=1mm,         
        right=1mm,        
        top=1mm,          
        bottom=1mm,       
        width=\textwidth, 
        before skip=0.1pt,
        after skip=0.1pt,
    },
    context/.style={
        fontupper=\scriptsize,
        fonttitle=\large,
        colframe=querycol,     
        coltitle=white,   
        colback=white,    
        boxrule=0.3mm,    
        arc=2mm,          
        outer arc=2mm,    
        left=1mm,         
        right=1mm,        
        top=1mm,          
        bottom=1mm,       
        before skip=1pt,
        after skip=0.1pt, 
    },
    query/.style={
        fontupper=\scriptsize,
        fontlower=\scriptsize,
        colframe=querycol,     
        coltitle=white,   
        colback=white,    
        boxrule=0.1mm,    
        arc=2mm,          
        outer arc=2mm,    
        left=1mm,         
        right=1mm,        
        top=1mm,          
        bottom=1mm,       
        before skip=1pt,
        after skip=0.1pt,
    },
    abab/.style={
        fontupper=\scriptsize,
        fonttitle=,
        colframe=ababcol, 
        coltitle=white,   
        boxrule=0.5mm,    
        arc=2mm,          
        outer arc=2mm,    
        left=1mm,         
        right=1mm,        
        top=1mm,          
        bottom=1mm,       
        width=0.33\textwidth, 
        before skip=0.1pt,
        after skip=0.1pt, 
    },
    others/.style={
        fontupper=\scriptsize,
        colframe=myhailuo3light, 
        coltitle=white,
        boxrule=0.5mm,    
        arc=2mm,          
        outer arc=2mm,    
        left=1mm,         
        right=1mm,        
        top=1mm,          
        bottom=1mm,       
        width=0.33\textwidth, 
        before skip=0.1pt,
        after skip=0.1pt, 
    },
    goldanswer/.style={
        fontupper=\scriptsize,
        colframe=goldanswercol,     
        coltitle=white,   
        boxrule=0.5mm,    
        arc=2mm,          
        outer arc=2mm,    
        left=1mm,         
        right=1mm,        
        top=1mm,          
        bottom=1mm,       
        width=0.33\textwidth, 
        before skip=0.1pt,
        after skip=0.1pt, 
    },
}

\theoremstyle{plain}

\theoremstyle{definition}

\theoremstyle{remark}

\usepackage{CJKutf8}

\definecolor{medgray55}{gray}{0.55}
\definecolor{medgray}{gray}{0.7}
\definecolor{litegray}{gray}{0.9}
\definecolor{gblue}{RGB}{210, 227, 252}
\definecolor{gred}{RGB}{250, 210, 207}
\definecolor{gyellow}{RGB}{254, 239, 195}
\definecolor{ggreen}{RGB}{206, 234, 214}
\definecolor{gorange}{RGB}{254, 223, 200}

\definecolor{gblue9}{RGB}{23, 78, 166}
\definecolor{gred9}{RGB}{165, 14, 14}
\definecolor{gyellow9}{RGB}{227, 116, 0}
\definecolor{ggreen9}{RGB}{13, 101, 45}
\definecolor{gorange9}{RGB}{176, 96, 0}

\definecolor{myblue}{rgb}{0,0,1}
\definecolor{myred}{rgb}{1,0,0}
\definecolor{mylightgray}{gray}{0.95}

\definecolor{highlightblue}{HTML}{185ABC}

\makeatletter

\renewcommand\paragraph{\@startsection{paragraph}{4}{\z@}%
            {-2.5ex\@plus -1ex \@minus -.25ex}%
            {1.25ex \@plus .25ex}%
            {\itshape\normalsize\bfseries}}
\makeatother
\newcolumntype{L}[1]{>{\raggedright\let\newline\\\arraybackslash\hspace{0pt}}m{#1}}
\newcolumntype{C}[1]{>{\centering}m{#1}}

\newcolumntype{R}[1]{>{\raggedleft\let\newline\\\arraybackslash\hspace{0pt}}m{#1}}

\definecolor{ao}{rgb}{0.0, 0.0, 1.0}

\newcommand\vcent[1]{\vcenter{\hbox{#1}}}

\newcommand\loudspeaker[1][3]{\ensuremath{\vcent{\rule{.6ex}{.6ex}}\kern-.5ex%
  \vcent{\scalebox{.6}[1]{\rotatebox[origin=center]{90}{$\blacktriangle$}}}%
  \ifnum#1>0\relax\kern.05ex\vcent{\scalebox{.4}{\ttfamily)}}%
  \ifnum#1>1\relax\kern-.4ex\vcent{\scalebox{.56}{\ttfamily)}}%
  \ifnum#1>2\relax\kern-.55ex\vcent{\scalebox{.7}{\ttfamily)}}%
  \fi\fi\fi}%
}

\definecolor{green}{rgb}{0.9,0.9,0.9}

\makeatletter
\renewcommand\subparagraph{%
 \@startsection {subparagraph}{5}{\z@ }{3.25ex \@plus 1ex
 \@minus .2ex}{-1em}{\normalfont \normalsize \bfseries }}%
\makeatother

\let\cite\citep

\reportnumber{} 

\renewcommand{\today}{}

\author[*,1]{MiniMax\footnote{Please send correspondence to model@minimax.io.}}

\begin{abstract}
We introduce MiniMax-Speech, an autoregressive Transformer-based Text-to-Speech (TTS) model that generates high-quality speech. A key innovation is our learnable speaker encoder, which extracts timbre features from a reference audio without requiring its transcription. This enables MiniMax-Speech to produce highly expressive speech with timbre consistent with the reference in a zero-shot manner, while also supporting one-shot voice cloning with exceptionally high similarity to the reference voice. In addition, the overall quality of the synthesized audio is enhanced through the proposed Flow-VAE. Our model supports 32 languages and demonstrates excellent performance across multiple objective and subjective evaluations metrics. Notably, it achieves state-of-the-art (SOTA) results on objective voice cloning metrics (Word Error Rate and Speaker Similarity) and has secured the top position on the public TTS Arena leaderboard. Another key strength of MiniMax-Speech, granted by the robust and disentangled representations from the speaker encoder, is its extensibility without modifying the base model, enabling various applications such as: arbitrary voice emotion control via LoRA; text to voice (T2V) by synthesizing timbre features directly from text description; and professional voice cloning (PVC) by fine-tuning timbre features with additional data. We encourage readers to visit \href{https://minimax-ai.github.io/tts_tech_report}{https://minimax-ai.github.io/tts\_tech\_report} for more examples.
\end{abstract}

\begin{document}
\title{}
\maketitle

\begin{figure}[t]
\centering
  \includegraphics[width=\textwidth]{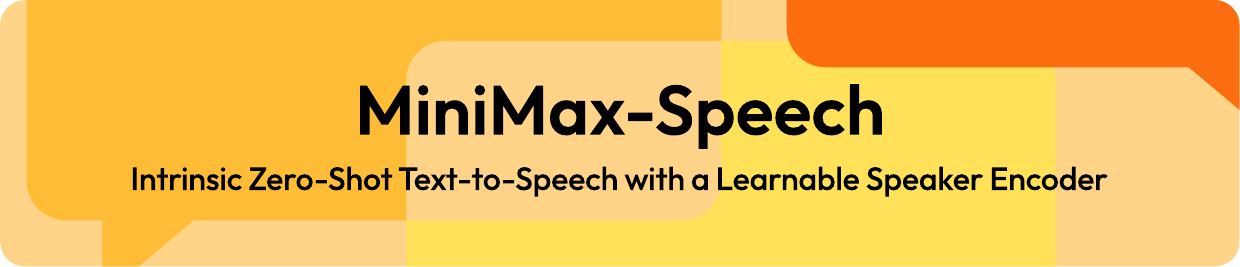}
  \label{fig:banner}
\end{figure}

\section{Introduction}
\label{sec:intro}
The advent of codec-based models has catalyzed significant advancements in TTS. These models, when trained on large-scale datasets, demonstrate the remarkable capability to generate high-quality speech from only a few seconds of reference audio. This  proficiency has fostered their widespread adoption across diverse applications, including conversational AI, audio content creation for blogs, interactive voice assistants, and immersive e-book narration.

Two primary modeling methodologies are prevalent in large TTS models: autoregressive (AR) language models and non-autoregressive (NAR) diffusion models. NAR diffusion models have gained attention for their rapid inference capabilities. However, such NAR models often employ duration modeling techniques. One approach is explicit phoneme-duration alignment~\cite{ju2024naturalspeech, le2023voicebox, mehta2024matcha, gao2023e3}, which can constrain naturalness and diversity. An alternative, global duration modeling~\cite{wang2024maskgct, anastassiou2024seed, lee2024ditto, yang2024simplespeech, chen2024f5, eskimez2024e2, jiang2025megatts}, involves the model learning implicit alignments; this, however, can increase modeling complexity and reduce robustness in challenging cases. Conversely, AR models are renowned for their potent modeling capacity, an inherent strength that allows them to generate speech exhibiting superior prosody, intonation, and overall naturalness~\cite{borsos2023audiolm, wang2023neural, casanova2024xtts, wang2025spark, guo2024fireredtts}.

Most previous AR TTS models~\cite{wang2023neural, du2024cosyvoice2} have required both speech and transcription as prompts in voice cloning, a methodology categorized as one-shot learning. However, semantic or linguistic mismatches between prompt and target speech, compounded by decoding length limitations, often result in suboptimal generation quality. MiniMax-Speech distinguishes itself by integrating a speaker encoder into its AR model. This key feature, based on concepts explored in works like~\cite{betker2023better}, enables zero-shot voice cloning using only a speech prompt. This approach eliminates the dependency on reference transcription, thus inherently supporting cross-lingual and multilingual synthesis and avoiding issues from text-speech mismatches. By conditioning solely on vocal characteristics, it allows for a wider, more flexible decoding space for prosody and style during generation, leading to richer and more varied outputs. Notably, even in one-shot scenarios where a reference transcription is available, this integrated speaker encoder contributes to superior speaker similarity in the synthesized speech. The strategy employed by MiniMax-Speech thus effectively mitigates these common issues and offers greater flexibility for various extensions, such as T2V and PVC.

Another technical aspect warranting discussion is the inherent learnability of the speaker encoder we adopted~\cite{casanova2024xtts, betker2023better}. Some models often utilize modules pre-trained on Speaker Verification (SV) tasks as their fixed speaker encoders~\cite{du2024cosyvoice}. However, the training data and optimization objectives inherent to SV tasks can differ from the specific requirements of the TTS task itself. In contrast, our learnable approach involves jointly training the speaker encoder with the AR model. This end-to-end optimization strategy allows the speaker encoder to be better tailored to the demands of the TTS task. Subsequent experimental results compellingly demonstrate that employing a learnable speaker encoder yields superior performance in terms of both speaker similarity and intelligibility of the synthesized speech.

Flow matching models~\cite{lipman2022flow, tong2023improving}, when utilized as decoders, are capable of producing high-fidelity speech outputs~\cite{du2024cosyvoice, du2024cosyvoice2}. A prevalent paradigm involves flow matching models first predicting mel-spectrogram, which is subsequently converted to an audio waveform by a vocoder. However, the mel-spectrogram can act as an information bottleneck, inherently limiting the ultimate achievable speech quality. In contrast, VAE, benefiting from end-to-end training, demonstrates stronger representation learning capability. Employing VAE-derived representations as the modeling objective for flow matching systems subsequently improves speech quality~\cite{tan2024naturalspeech, kim2021conditional, hung2024tangoflux}. Building upon this VAE foundation, MiniMax-Speech innovatively introduces Flow-VAE. This hybrid approach integrates VAE and flow models to enhance the information representation power of the VAE encoder, thereby further improving both audio quality and speaker similarity.

\vspace{5pt}
We summarize our contributions as follows:
\vspace{-5pt}
\begin{enumerate}
\item We present MiniMax-Speech, a TTS model that supports 32 languages and generates high-fidelity speech with near-indistinguishable human resemblance, achieving SOTA results on multiple objective and subjective evaluation metrics.

\item Based on autoregressive Transformer architecture and equipped with a learnable speaker encoder module, our model demonstrates strong expressiveness in zero-shot voice cloning. Furthermore, it enhances speaker similarity in one-shot scenarios when provided with a reference audio prompt.

\item We employ a flow matching model based on our novel Flow-VAE, which further improves the audio quality and speaker similarity of the generated speech. 

\item We also detail several downstream applications of our model,  including fine-grained control over emotional expression in synthesized speech, extensive voice library via T2V, and improving synthesis similarity for target speakers by PVC.

\end{enumerate}

\section{Method}

\begin{figure}[t]
\centering
  \includegraphics[width=\textwidth]{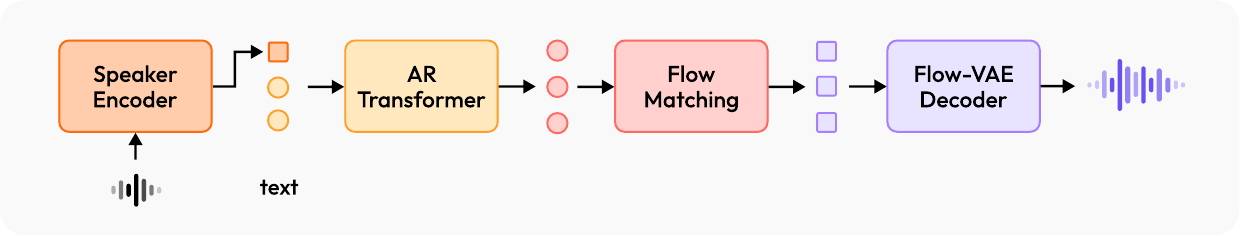}
  \caption{\textbf{An overview of the architecture of MiniMax-Speech.}}
  \label{fig:overview}
\end{figure}

MiniMax-Speech is an innovative TTS system designed for high-fidelity voice cloning, with a particular emphasis on its robust zero-shot capabilities. As shown in Figure~\ref{fig:overview}, it primarily comprises three components: a tokenizer, an autoregressive Transformer, and a latent flow matching model, which consists of flow matching module and Flow-VAE module. The text tokenizer utilizes Byte Pair Encoding (BPE), while the audio tokenizer employs Encoder-VQ-Decoder architecture~\cite{van2017neural, betker2023better} quantization on mel-spectrograms with a rate of 25 tokens per second and connectionist temporal classification (CTC) supervision. This speech tokenizer achieves a high compression rate, while effectively preserving ample acoustic details and semantic information. The details of the autoregressive Transformer and latent flow matching model are as follows.

\subsection{Zero-Shot Autoregressive Transformer}

MiniMax-Speech employs an autoregressive Transformer~\cite{vaswani2017attention} architecture to generate discrete audio tokens from textual input. The system excels at high-fidelity speaker cloning, particularly zero-shot voice cloning, where it synthesizes speech emulating a target speaker's distinct timbre and style from only a single, untranscribed audio segment.

To enable this powerful zero-shot capability, MiniMax-Speech incorporates a learnable speaker encoder, inspired by~\cite{betker2023better}. In contrast to other speech synthesis models that use pre-trained speaker encoders~\cite{lajszczak2024base, du2024cosyvoice}, the encoder in MiniMax-Speech is trained jointly with the autoregressive Transformer. This joint optimization allows the speaker encoder to be specifically tailored to the speech synthesis task, enhancing the model's synthesis quality by providing richer and more relevant speaker-specific information. Additionally, since the speaker encoder is learnable, it can be trained on all languages within the training dataset. Compared to pre-trained speaker encoders that might not have been exposed to the same diversity of languages, our learnable speaker encoder ensures broader linguistic coverage and potentially enhances generalization.

The speaker encoder extracts salient speaker-specific characteristics, such as vocal timbre and prosodic style, from the reference audio (which differs from the target speech to be generated). Variable-length audio segments serving as voice prompts are transformed by this encoder into a fixed-size conditional vector. This vector subsequently guides the autoregressive model in generating the target speech with the desired speaker identity.

\begin{figure}[t]
\centering
  \includegraphics[width=\textwidth]{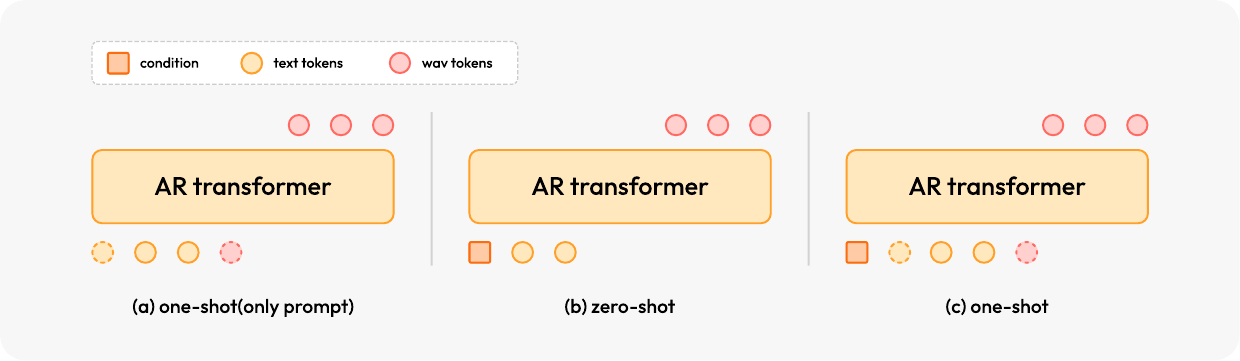}
  \caption{\textbf{Different Voice Cloning Approaches in AR Transformer.} The dotted line represents a provided example of a text-to-speech pair.}
  \label{fig:GPT}
\end{figure}

The voice cloning capabilities of Minimax-Speech are best understood through the paradigms of zero-shot and one-shot learning, concepts adapted from the capabilities observed in Large Language Models (LLMs) like GPT-3~\cite{brown2020language}. In LLMs, zero-shot refers to performing a task based solely on an instruction without any prior examples, while one-shot (or few-shot) involves providing one (or a few) in-context examples to guide the model. We adapt these concepts to TTS as follows:

\begin{itemize}
\item \textbf{Zero-Shot Voice Cloning}: The Core Strength of MiniMax-Speech. In this primary mode, Minimax-Speech synthesizes speech in a target speaker's voice using only a reference audio segment to define the voice characteristics (shown in Figure~\ref{fig:GPT}b). Crucially, no explicit examples of that speaker's voice paired with text are provided at inference time as prompts, and no speaker-specific fine-tuning is performed. The reference audio itself acts as the primary "instruction" for the desired vocal timbre and style.

\item \textbf{One-Shot Voice Cloning}: An Optional Enhancement. Building upon the zero-shot foundation, this mode enhances cloning fidelity by providing an additional explicit example. Specifically, a paired text-audio sample from the target speaker is supplied as an "in-context" prompt alongside the standard speaker embedding derived from the reference audio (shown in Figure~\ref{fig:GPT}c). This approach mirrors the one-shot prompting strategy in LLMs and is similar to techniques in prior works like VALL-E~\cite{wang2023neural}, CosyVoice 2~\cite{du2024cosyvoice2} and Seed-TTS~\cite{anastassiou2024seed} (their prompting method, which requires a paired text-audio sample, is illustrated in Figure~\ref{fig:GPT}a). While these aforementioned models are often described as "zero-shot" in their respective publications, their reliance on a paired text-audio prompt for speaker conditioning categorizes them as "one-shot" methods according to our stricter definition. Our "instrinsic zero-shot" approach (Figure~\ref{fig:GPT}b), in contrast, exclusively utilizes an untranscribed reference audio segment to derive speaker characteristics, without any accompanying text prompt.

\end{itemize}

While the optional one-shot prompting can offer finer-grained stylistic cues in specific scenarios, the system's architecture is fundamentally designed for powerful and flexible zero-shot synthesis. The conditioning encoder in MiniMax-Speech seamlessly supports both methods, but its true innovation lies in enabling high-quality voice cloning without reliance on paired data or fine-tuning. The advantages of this zero-shot-centric design, facilitated by the learnable speaker encoder, are manifold:

\begin{itemize}
\item \textbf{Text-Free Reference}: By operating solely on the reference audio waveform, it eliminates the need for textual transcriptions of the target speaker's audio. This ensures that the speaker identity is learned purely from vocal characteristics, disentangled from the semantic content of any specific reference utterance.

\item \textbf{Rich Prosodic Variation and Flexible Decoding}: The zero-shot approach, conditioned only on the speaker condition extracted by the encoder, allows for the generation of speech with diverse prosodic variations. The model is not constrained by the prosody of a specific text-audio prompt (as in one-shot methods), leading to a wider decoding space and outputs that maintain high fidelity to the target speaker's unique vocal identity while exhibiting a natural range of expression.

\item \textbf{Robust Cross-Lingual Synthesis}: The speaker encoder captures language-agnostic vocal characteristics, enhancing cross-lingual synthesis. This outperforms prompt-based cloning methods that rely on text-speech reference pairs, which struggle when the reference language differs from the target language or when semantic content mismatches.

\item \textbf{Foundation for Extensibility}: The robust and disentangled speaker representation provided by the encoder serves as a flexible foundation for various downstream applications, as detailed in Section~\ref{sec:extensions}. Tasks like emotion control, T2V, and PVC can leverage this core speaker identity representation without fundamentally altering the base model.
\end{itemize}

\subsection{Latent Flow Matching}

\subsubsection{Overview}

\begin{figure}[t]
    \centering
     \includegraphics[width=\textwidth]{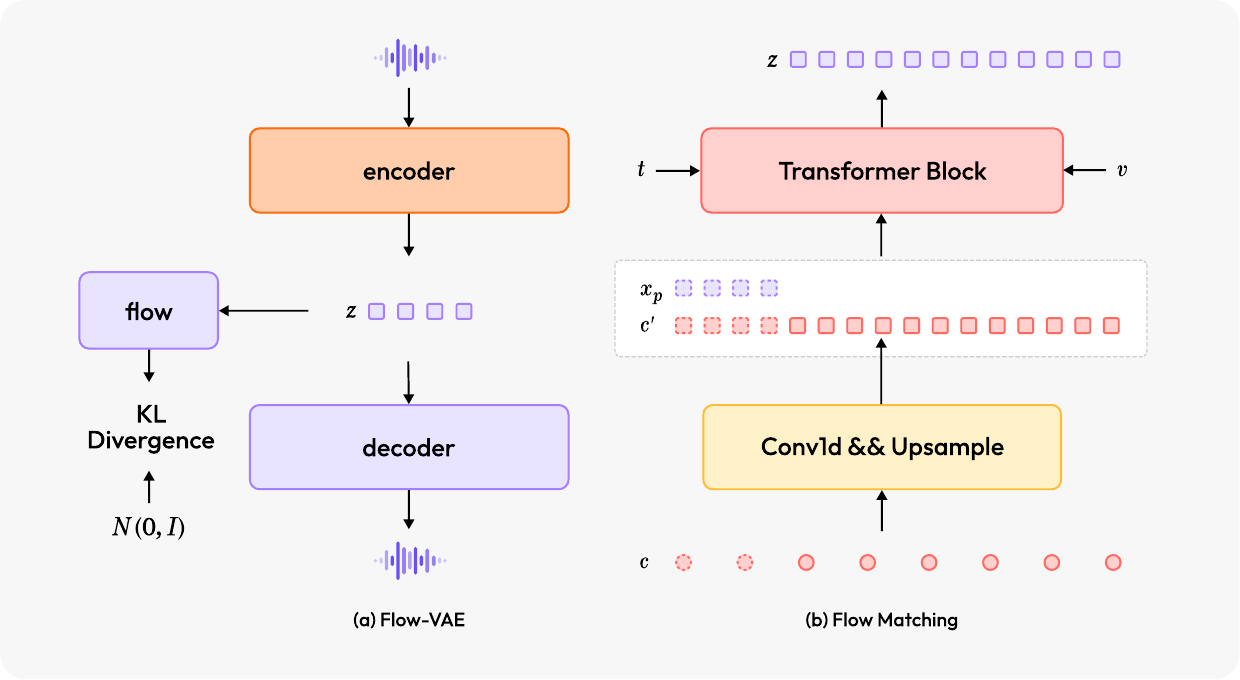}
    \caption{\textbf{An overview of the proposed latent flow matching Architecture.} (a) The Flow-VAE model consists of an encoder, which is used to extract continuous speech features $z$, and a decoder, which is used to restore continuous speech features back to waveforms, and a flow model, which converts the distribution of continuous speech features to a standard normal distribution. $z$ represents continuous speech features, which is the target for the flow matching model to generate. (b) The flow matching model, which conditions on AR transformer output $c$, a speaker embedding $v$, provided continuous speech features $x_p$ and intermediate state $x_t$ at timestep $t$ on the probabilistic density path. $c'$ represents the output of AR transformer after upsampling.  The dotted box indicates the provided prompt information.}
    \label{fig:flowvae_flowmatching}
\end{figure}

In MiniMax-Speech, the flow matching model utilizes a transformer~\cite{vaswani2017attention} architecture, which has powerful context modeling capabilities. Our flow matching model is designed to model the distribution of continuous speech features (latent), which are extracted from an encoder-decoder module trained on audio, rather than mel-spectrograms. When training this encoder-decoder module (where its encoder extracts these continuous speech features and its decoder is typically a neural vocoder~\cite{oord2016wavenet,kalchbrenner2018efficient,valin2019lpcnet,yang2021multi,kumar2019melgan}), the KL divergence is employed as a constraint. This renders the latent distribution easier to predict and more compact. Furthermore, due to the joint training of the latent extraction module (encoder) and the neural vocoder (decoder), the waveform reconstruction error from latent features is smaller compared to that from mel-spectrograms, which elevates the ceiling of latent feature modeling.

Figure~\ref{fig:flowvae_flowmatching}(a) illustrates the proposed Flow-VAE model, which we use to optimize the encoder-decoder module. Traditional Variational Autoencoders (VAEs) typically assume a standard normal distribution for their latent space. In contrast, Flow-VAE introduces a flow model~\cite{rezende2015variational,dinh2014nice,dinh2016density,kingma2018glow}, which can flexibly transform the latent space using a series of reversible mappings to learn a more expressive posterior distribution to more accurately capture complex patterns in the data. This fusion solution can make full use of VAE's initial modeling ability of data and the flow model's accurate fitting ability of complex distribution, so as to better capture the complex structure and distribution characteristics in the data, improve the accuracy of data modeling, and thus significantly outperforming the traditional VAE model.

To enhance the audio quality and timbre similarity of the flow matching model, inspired by CosyVoice 2~\cite{du2024cosyvoice2}, we incorporate both global timbre information and prompt information as mentioned in Figure~\ref{fig:flowvae_flowmatching}(b). Specifically, the global timbre information is extracted from mel-spectrogram features using the speaker encoder. During the training process, information from the beginning of the current sentence is utilized as a prompt with a certain probability. Consequently, at inference stage, our model supports both zero-shot and one-shot synthesis modalities.

\subsubsection{KL-Divergence for Flow-VAE}

In the Flow-VAE model, our goal is to provide enough information for the posterior encoder, which is the encoder of the Flow-VAE model, so we use the waveform of the target speech $x$ as input instead of the mel spectrogram, then we apply a flow model $f_\theta$ to reversibly transform a normal distribution into a standard normal distribution. The KL divergence is:

  \begin{equation}
    \begin{array}{c}
     L_{kl} = D_{KL}(q_\phi(\tilde{z} |x)||p(\tilde{z} )) = \log{q_\phi } (\tilde{z} |x) - \log{p(\tilde{z} )}
    \end{array}
  \label{eq1}
  \end{equation}
  
  \begin{equation}
    \begin{array}{c}
     q_{\phi}(\tilde{z} |x)=N(f_\theta(\tilde{z} ); \mu_{\phi}(x), \sigma_{\phi}(x))\left | det\frac{\partial f_\theta (\tilde{z} )}{\partial \tilde{z} }  \right | 
    \end{array}
  \label{eq2}
  \end{equation}

  \begin{equation}
    z ~\sim N(\mu_{\phi}(x), \sigma_{\phi}(x))
  \label{eq3}
  \end{equation}
  
  \begin{equation}
    p(\tilde{z} ) = N(\tilde{z} , 0, I)
  \label{eq4}
  \end{equation}

In our experiment, as show in Figure~\ref{fig:flowvae_flowmatching}(a), the flow model transforms the normal distribution output by the encoder through a series of reversible transformations. Finally, we calculate the KL loss between the distribution output by the flow model and the standard normal distribution. In this way, the output of the encoder can be constrained to a normal distribution instead of a standard normal distribution, enhancing the information expression ability of the encoder.

\section{Experiments}
This section presents a comprehensive evaluation of MiniMax-Speech, systematically assessing its performance across multiple dimensions. We begin by describing the datasets employed for training and evaluation. Our main analysis focuses on three key aspects: (1) voice cloning fidelity, objectively measured for both zero-shot and one-shot approaches; (2) perceptual naturalness, evaluated through extensive human preference tests; and (3) multilingual and cross-lingual synthesis capabilities, rigorously tested across diverse languages. Additionally, we conduct ablation studies to investigate the impact of key architectural decisions, including speaker conditioning methodology and the Flow-VAE framework.

\subsection{Datasets}
Minimax-Speech is trained on a multilingual speech dataset spanning 32 languages. Throughout the collection process, recognizing the paramount importance of transcription accuracy, we implemented a rigorous dual Automatic Speech Recognition (ASR) verification process. Text punctuation was further refined through the comprehensive consideration of VAD and ASR-generated timestamps. Notably, the original steady-state noise inherent in the recordings was preserved. Furthermore, consistent vocal timbre was maintained within each audio file by a multi-speaker verification model.

\subsection{Voice Clone Evaluation}
\label{sec:cloning_eval}

The fidelity of voice cloning was quantitatively assessed using WER and SIM metrics on the Seed-TTS-eval~\cite{anastassiou2024seed} test set. This dataset comprises two distinct subsets: test-zh (approximately 2,000 Chinese samples) and test-en (approximately 1,000 English samples). Each sample in these subsets includes a reference audio and a corresponding ground-truth audio from the identical speaker. For WER computation, synthesized English and Chinese audio were transcribed using Whisper-large-v3~\cite{radford2023robust} and Paraformer-zh~\cite{gao2023funasr}, respectively. SIM was determined by calculating the cosine similarity between speaker embeddings, which were extracted using a speaker verification model fine-tuned on WavLM-large. These choices of ASR and speaker verification models align with the established methodology of the Seed-TTS-eval test set.


\begin{table}[htbp]
\centering
\caption{\textbf{Objective Evaluation Metrics on the Seed-TTS Test Set.} The \textbf{bolded value} indicates the best indicator for each column, and the \underline{underlined value} indicates the second best. Note that the reference audio is utilized as input to the speaker encoder in both cloning methods of MiniMax-Speech, and it additionally serves as prompt exemplar for the one-shot paradigm.}
\label{tab:objective_metrics_seedtts}
\begin{tabular}{lccccc}
\toprule
&                       & \multicolumn{2}{c}{test-zh}    & \multicolumn{2}{c}{test-en}    \\ \cline{3-6} 
\multirow{-2}{*}{Model} & \multirow{-2}{*}{Cloning Method} & WER $\downarrow$          & SIM $\uparrow$            & WER $\downarrow$           & SIM $\uparrow$           \\ \hline
Ground Truth            & -                                & 1.25          & 0.750          & 2.14          & 0.730          \\
Seed-TTS                & one-shot                         & 1.12          & \underline{0.796}          & 2.25          & \textbf{0.762} \\
CosyVoice 2              & one-shot                         & 1.45          & 0.748          &  2.57          &  0.652          \\
\rowcolor[HTML]{eee6ff} 
MiniMax-Speech          & zero-shot                        & \textbf{0.83} & 0.783          & \textbf{1.65} & 0.692          \\
\rowcolor[HTML]{eee6ff} 
MiniMax-Speech          & one-shot                         & \underline{0.99}          & \textbf{0.799} & \underline{1.90}          & \underline{0.738}          \\ 
\bottomrule
\end{tabular}
\end{table}

As presented in Table~\ref{tab:objective_metrics_seedtts}, the MiniMax-Speech model achieved markedly lower WER in both zero-shot and one-shot cloning scenarios compared to Seed-TTS~\cite{anastassiou2024seed}, CosyVoice 2~\cite{du2024cosyvoice2}, and ground truth audio. This demonstrates that speech synthesized by MiniMax-Speech during cloning is characterized by clear, stable pronunciation and a reduced incidence of articulation errors. Notably, the WER for MiniMax-Speech in zero-shot cloning was superior to its one-shot counterpart. Furthermore, subjective listener feedback indicated that speech synthesized via zero-shot cloning was perceived as more natural and realistic. The zero-shot approach, empowered by our speaker encoder, directly leverages the reference audio's acoustic properties without the additional influence of a language model prompt exemplar. This leads to superior intelligibility (lower WER) and enhanced naturalness, as the model has greater freedom in generating prosody faithful to the text being synthesized, rather than being biased by the prompt's prosody. The speaker encoder effectively captures the core vocal identity, allowing the autoregressive model to generate diverse and natural speech. While one-shot prompting improves SIM, the zero-shot method demonstrates a compelling balance of clarity and naturalness.

Regarding SIM, the MiniMax-Speech model achieved a SIM score in zero-shot cloning comparable to that of the ground truth audio. This underscores the speaker encoder's efficacy in extracting and preserving speaker identity even without textual or prosodic cues from a prompt. When an exemplar audio was introduced as a prompt in the one-shot cloning setting, the SIM score surpassed that of the ground truth audio, outperformed CosyVoice2, and was on par with Seed-TTS. This finding suggests that the incorporation of a prompt exemplar, building upon the zero-shot approach, can further augment the similarity of the cloned voice, potentially by providing more explicit cues for fine-grained vocal characteristics.

\subsection{Subjective Evaluation}
\label{sec:subjective_eval}

To comprehensively evaluate MiniMax-Speech in real-world scenarios, we submitted our model to Artificial Arena\footnote{Artificial Arena: \url{https://artificialanalysis.ai/text-to-speech/arena?tab=leaderboard}}, a public TTS model leaderboard. Artificial Arena ranks models using ELO scores, derived from human preference judgments as users listen to and compare speech samples from various models. For this demanding evaluation, all speech samples from MiniMax-Speech were generated using its advanced zero-shot speaker cloning capability. This approach, while offering immense flexibility, presents a significant challenge in achieving SOTA quality.

\begin{figure}[t]
\centering
\includegraphics[scale=0.28]{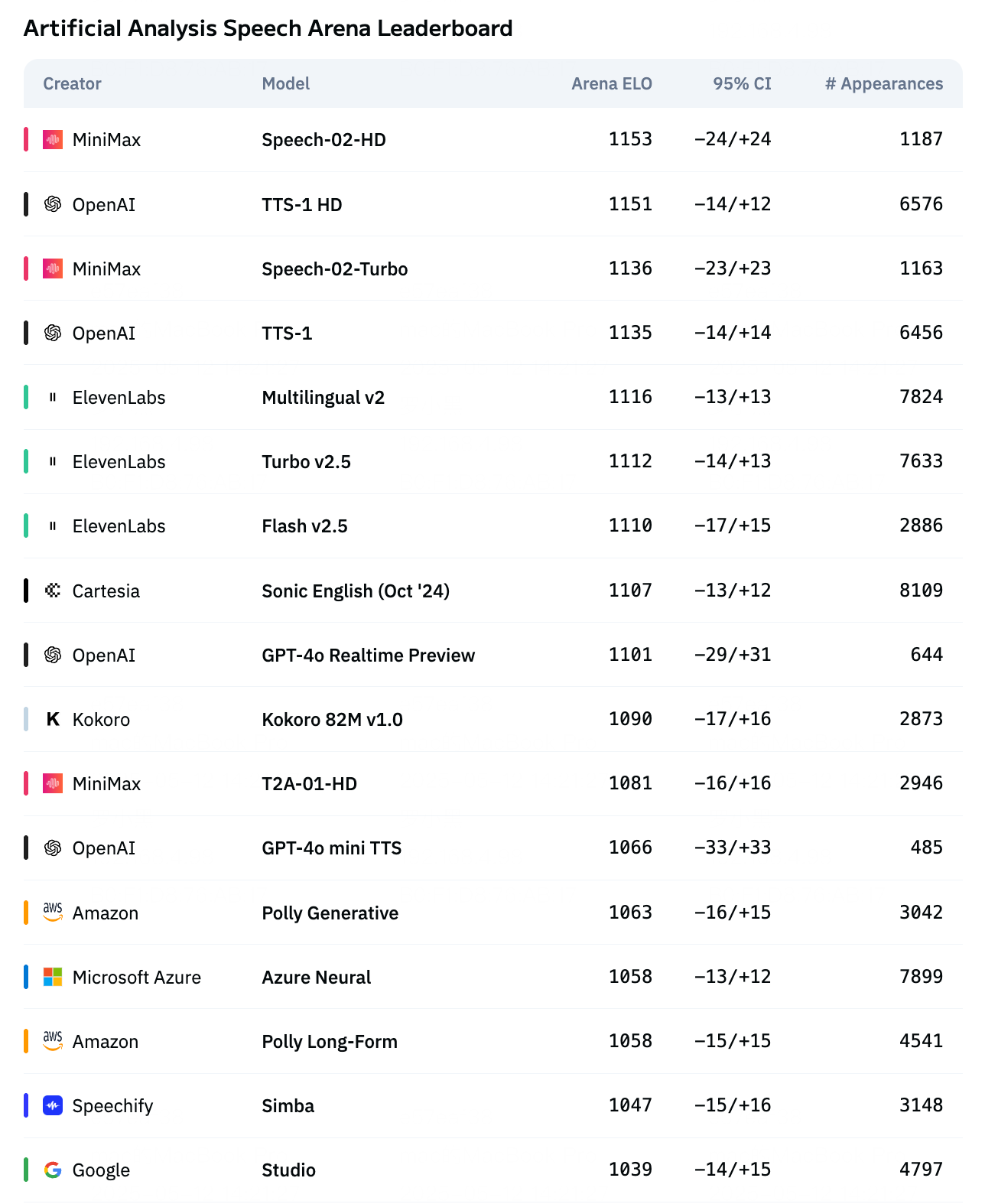}
\caption{\textbf{Artificial Arena Evaluation Results (May 12, 2025).}}
\label{fig:arena_results}
\end{figure}

As presented in Figure~\ref{fig:arena_results}, our model, MiniMax-Speech (referred to as Speech-02-HD\footnote{The MiniMax-Speech model described in this paper corresponds to the Speech-02-HD model in Artificial Arena. The Speech-02-Turbo model, which also participated in the evaluation, employs a different model architecture primarily to enhance inference speed and reduce operational costs. The T2A-01-HD model is an older version of our model.} on the leaderboard), secured the leading position. This top ranking not only places MiniMax-Speech ahead of a strong field of competitors but also underscores its distinct advantages. Specifically, when compared to other leading models such as those from OpenAI and ElevenLabs, the ELO scores reflect a clear user preference for MiniMax-Speech's superior naturalness and heightened expressiveness.

Perhaps even more strikingly, MiniMax-Speech demonstrated a considerable ELO score advantage over models from major technology providers like Google, Microsoft, and Amazon. This significant gap suggests that MiniMax-Speech's underlying architecture represents a more advanced, next-generation approach. Critically, our zero-shot generated speech achieved a level of quality and user preference that surpassed systems which often rely on extensive data to individually train models for specific speakers (e.g., requiring tens of hours of audio per voice) to reach their peak performance.

Crucially, achieving this high degree of naturalness and expressiveness—sufficient to outperform leading industry models, including those potentially built on extensive speaker-specific data—while relying exclusively on zero-shot cloned speaker timbres, robustly underscores the advanced capabilities and generalization power of our model. This outstanding performance in a public, preference-based benchmark highlights MiniMax-Speech's ability to deliver a highly compelling and human-like listening experience in real-world applications, even when generating novel voices on the fly.

\subsection{Multilingual Evaluation}

\label{sec:multilingual_eval}
MiniMax-Speech supports synthesis in 32 languages. To assess its multilingual performance, we constructed a dedicated test set comprising 24 languages\footnote{\url{https://huggingface.co/datasets/MiniMaxAI/TTS-Multilingual-Test-Set}}. For each language, this set includes 100 distinct test sentences. Synthesized speech was generated using the cloned voices of two speakers per language—one male and one female—sourced from the Mozilla Common Voice dataset~\cite{ardila2019common}. Each speaker rendered 50 unique sentences from the 100 available for that language. We assessed the performance of MiniMax-Speech against the ElevenLabs Multilingual v2 model in multilingual synthesis. Speech from both models was synthesized using zero-shot voice cloning. The evaluation metrics and methodology are consistent with those described in Section \ref{sec:cloning_eval}. For all languages except Chinese, the Whisper-large-v3 model was used for text recognition.

\begin{table}[h!]
\centering
\caption{\textbf{Objective Evaluation Metrics on the Multilingual Test Set.}}
\label{tab:multilingual_results}
\resizebox{0.65\textwidth}{!}{%
\begin{tabular}{lcccc}
\toprule
Language & \multicolumn{2}{c}{WER $\downarrow$} & \multicolumn{2}{c}{SIM $\uparrow$} \\
\cmidrule(lr){2-3} \cmidrule(lr){4-5}
 & MiniMax & ElevenLabs & MiniMax & ElevenLabs \\
\midrule
Chinese    & \textbf{2.252} & 16.026         & \textbf{0.780} & 0.677          \\
English    & \textbf{2.164} & 2.339          & \textbf{0.756} & 0.613          \\
Cantonese  & \textbf{34.111}& 51.513         & \textbf{0.778} & 0.670          \\
Japanese   & \textbf{3.519} & 10.646         & \textbf{0.776} & 0.738          \\
Korean     & \textbf{1.747} & 1.865          & \textbf{0.776} & 0.700          \\
Arabic     & \textbf{1.665} & 1.666          & \textbf{0.736} & 0.706          \\
Spanish    & \textbf{1.029} & 1.084          & \textbf{0.762} & 0.615          \\
French     & \textbf{4.099} & 5.216          & \textbf{0.628} & 0.535          \\
Italian    & \textbf{1.543} & 1.743          & \textbf{0.699} & 0.579          \\
Vietnamese & \textbf{0.880} & 73.415         & \textbf{0.743} & 0.369          \\
Thai       & \textbf{2.701} & 73.936         & \textbf{0.800} & 0.588          \\
Turkish    & 1.520          & \textbf{0.699} & \textbf{0.779} & 0.596          \\
Indonesian & 1.237          & \textbf{1.059} & \textbf{0.729} & 0.660          \\
Portuguese & 1.877          & \textbf{1.331} & \textbf{0.805} & 0.711          \\
Dutch      & 1.143          & \textbf{0.803} & \textbf{0.738} & 0.680          \\
German     & 1.906          & \textbf{0.572} & \textbf{0.733} & 0.614          \\
Russian    & 4.281          & \textbf{3.878} & \textbf{0.761} & 0.676          \\
Ukrainian  & 1.082          & \textbf{0.997} & \textbf{0.730} & 0.647          \\
Polish     & 1.415          & \textbf{0.766} & \textbf{0.802} & 0.729          \\
Romanian   & 2.878          & \textbf{1.347} & \textbf{0.809} & 0.699          \\
Greek      & 2.016          & \textbf{0.991} & \textbf{0.826} & 0.733          \\
Czech      & 3.875          & \textbf{2.108} & \textbf{0.796} & 0.685          \\
Finnish    & 4.666          & \textbf{2.964} & \textbf{0.835} & 0.759          \\
Hindi      & 6.962          & \textbf{5.827} & \textbf{0.818} & 0.730          \\
\bottomrule
\end{tabular}
}
\end{table}

As indicated in Table \ref{tab:multilingual_results}, regarding WER, the performance of MiniMax-Speech is comparable to that of Multilingual v2. For languages such as Chinese, Cantonese, Thai, Vietnamese, and Japanese, where Multilingual v2 exhibited a WER exceeding 10\%, MiniMax-Speech consistently outperformed it. This robust performance, particularly in languages with complex tonal structures or diverse phonetic inventories (e.g., Chinese, Cantonese, Thai, Vietnamese), suggests MiniMax-Speech's architecture is adept at capturing and reproducing nuanced acoustic features critical for intelligibility in these languages, an area where Multilingual v2 appears to face greater challenges. Concerning SIM, MiniMax-Speech demonstrated markedly superior SIM scores across all tested languages compared to the Multilingual v2 model. This consistent superiority in SIM across a diverse linguistic landscape underscores the effectiveness of MiniMax-Speech's speaker encoder and synthesis pipeline in preserving speaker identity, irrespective of the target language's phonetic characteristics, a key benefit of its text-agnostic reference processing. This suggests that MiniMax-Speech produces cloned speech that is closer to the ground truth human voice across the 24 evaluated languages.

\subsection{Cross-lingual Evaluation}
\label{sec:Cross-lingual_Synthesis}

A significant advantage of MiniMax-Speech, stemming from its speaker encoder architecture, is its inherent support for cross-lingual speech synthesis. This enables the synthesis of speech for any given speaker in all languages supported by the model. Two key aspects contribute to this capability:

Firstly, for zero-shot speaker cloning, MiniMax-Speech requires only a short audio segment from the target speaker, without any corresponding transcription. This minimalist data requirement significantly lowers the barrier to entry and operational complexity for cloning new voices. This contrasts with some one-shot cloning models~\cite{du2024cosyvoice2, anastassiou2024seed} that necessitate transcribed reference audio. Such a dependency on transcriptions not only complicates the cloning process but also introduces the risk of transcription errors negatively impacting the quality of the cloned voice. MiniMax-Speech's approach, by eliminating the need for transcriptions in its zero-shot cloning, simplifies the workflow and mitigates potential issues arising from inaccurate transcriptions.

Secondly, the speaker encoder module extracts a conditional vector that primarily captures voice timbre while being largely devoid of textual semantic information. This characteristic facilitates the model's ability to decouple voice timbre from linguistic content and subsequently recombine them, thereby enabling each distinct voice timbre to articulate speech across all supported languages.

To validate the cross-lingual synthesis capabilities enabled by the speaker encoder, we conducted evaluations using Chinese speakers from our multilingual test set. This involved synthesizing speech from these Chinese speakers uttering phrases in various other target languages.

\begin{table}[h!]
\centering
\caption{\textbf{Cross-lingual Speech Synthesis Performance of MiniMax-Speech (Zero-shot vs. One-shot).}}
\label{tab:crosslingual_synthesis}
\resizebox{0.65\textwidth}{!}{%
\begin{tabular}{lcccc}
\toprule
Target Language & \multicolumn{2}{c}{WER $\downarrow$} & \multicolumn{2}{c}{SIM $\uparrow$} \\
\cmidrule(lr){2-3} \cmidrule(lr){4-5}
 & Zero-Shot & One-Shot & Zero-Shot & One-Shot \\
\midrule
Czech      & \textbf{2.823} & 5.096          & 0.605          & \textbf{0.648} \\
Romanian   & \textbf{3.081} & 5.353          & 0.625          & \textbf{0.69}  \\
Finnish    & \textbf{4.527} & 8.112          & 0.554          & \textbf{0.655} \\
Thai       & \textbf{2.826} & 4.107          & 0.729          & \textbf{0.748} \\
Arabic     & \textbf{1.446} & 2.649          & 0.619          & \textbf{0.632} \\
French     & \textbf{4.497} & 5.489          & 0.586          & \textbf{0.645} \\
Vietnamese & \textbf{0.659} & 1.788          & 0.692          & \textbf{0.725} \\
\bottomrule
\end{tabular}
}
\end{table}

As indicated in Table~\ref{tab:crosslingual_synthesis}, MiniMax-Speech, when utilizing its zero-shot cloning method, achieves significantly lower WER across all tested languages compared to its one-shot method. Furthermore, the achieved WERs indicate a high level of intelligibility, approaching that of high-quality native synthesis in the target languages. These results demonstrate that the speaker encoder architecture provides MiniMax-Speech with excellent cross-lingual synthesis capabilities.

In contrast, while MiniMax-Speech's one-shot cloning approach yields higher SIM, its pronunciation accuracy in cross-lingual synthesis, as indicated by its notably higher WER, is considerably poorer. Consequently, these findings underscore the advantages of MiniMax-Speech's speaker encoder architecture, highlighting its flexibility in supporting both zero-shot and one-shot cloning paradigms, and particularly its superior performance in zero-shot cross-lingual synthesis, as evidenced by high pronunciation accuracy.

\subsection{Speaker Condition Evaluation}

To evaluate the effectiveness of different speaker conditioning approaches, we conducted an ablation study using three distinct models trained on a substantial subset of our Chinese speech data. The first model implemented our learnable speaker encoder architecture, the second utilized speaker embeddings (SpkEmbed) extracted from a pre-trained speaker verification model~\cite{cam++}, and the third employed a one-shot learning strategy with only an example audio prompt. We assessed these configurations using WER and SIM metrics.

\begin{table}[h!]
\centering
\caption{\textbf{Ablation Study on Speaker Conditioning Methods.}}
\label{tab:ablation_spk_condition}
\begin{tabular}{llcc}
\toprule
Method & Cloning Mode & WER $\downarrow$ & SIM $\uparrow$ \\
\midrule
Speaker Encoder  & Zero-shot & 1.252 & 0.730 \\
Speaker Encoder  & One-shot & 1.243 & 0.746 \\
SpkEmbed & Zero-shot & 1.400 & 0.746 \\
SpkEmbed & One-shot & 1.704 & 0.744 \\
OnlyPrompt & One-shot & 1.207 & 0.726 \\
\bottomrule
\end{tabular}
\end{table}

Analysis of Table~\ref{tab:ablation_spk_condition} demonstrates that the speaker encoder method provides the most robust performance, achieving strong results for both WER and SIM. In comparison, utilizing speaker embeddings from a pre-trained speaker model (SpkEmbed), while maintaining reasonable SIM, adversely affects the WER (e.g., 1.400 for SpkEmbed vs. 1.252 for Speaker Encoder in zero-shot), indicating a potential loss of speech clarity. This suggests an advantage of our learnable speaker encoder, which can be optimized jointly with the synthesis model, potentially adapting more effectively to the nuances of the target speech synthesis task compared to a fixed, pre-trained speaker verification model. Conversely, relying solely on the prompt (OnlyPrompt) in a one-shot setting, while achieving the best WER in this specific ablation study (1.207), significantly compromises SIM (0.726). 

Our learnable speaker encoder, especially in one-shot mode (WER 1.243, SIM 0.746), strikes an optimal balance, surpassing SpkEmbed in WER and OnlyPrompt in SIM. These results confirm its effectiveness in preserving both speech intelligibility and voice characteristics. It thus offers a more balanced speaker conditioning solution than the alternatives. Its ability to maintain strong speaker identity (SIM 0.730) and good intelligibility (WER 1.252) in zero-shot synthesis further underscores its advantages. However, the reference audio for the speaker encoder must differ from the target audio for AR Transformer synthesis. Using identical audio during training can cause semantic leakage and degrade performance.

\subsection{Flow-VAE Evaluation}

To evaluate the performance of VAE and Flow-VAE, we conducted comparisons in two primary aspects: vocoder resynthesis and TTS synthesis. We randomly selected a portion from the open-source Chinese and English test sets of Seed-TTS~\cite{anastassiou2024seed} as our test set.

\begin{table}[t]
\caption{\textbf{Objective indicators of resynthesis by VAE and Flow-VAE.} SELF-SIM represents the similarity between the synthesized audio and the original audio, and PROMPT-SIM represents the similarity between the synthesized audio and the prompt audio. }
\label{tab:objective_metrics_flowvae_vae}
\centering
\begin{tabular}{cccccccl}
\cline{1-7}
\multicolumn{1}{c}{Model}       & \multicolumn{1}{c}{SELF-SIM $\uparrow$}                      & \multicolumn{1}{c}{PROMPT-SIM $\uparrow$}                   & \multicolumn{1}{c}{NB PESQ $\uparrow$}                      & \multicolumn{1}{c}{WB PESQ $\uparrow$}                      & \multicolumn{1}{c}{STOI $\uparrow$}                          & \multicolumn{1}{c}{MS-STFT-LOSS $\downarrow$}                 & \multicolumn{1}{c}{} \\ \cline{1-7}
\multicolumn{1}{c}{Dac-Vae} & \multicolumn{1}{c}{0.98} & \multicolumn{1}{c}{0.748} & \multicolumn{1}{c}{4.27} & \multicolumn{1}{c}{4.20} & \multicolumn{1}{c}{0.993} & \multicolumn{1}{c}{0.67} &                      \\ \cline{1-7}

\multicolumn{1}{c}{\cellcolor[HTML]{eee6ff}Dac-Flow-Vae} & \multicolumn{1}{c}{\cellcolor[HTML]{eee6ff}\textbf{0.986}} & \multicolumn{1}{c}{\cellcolor[HTML]{eee6ff}\textbf{0.75}} & \multicolumn{1}{c}{\cellcolor[HTML]{eee6ff}\textbf{4.34}} & \multicolumn{1}{c}{\cellcolor[HTML]{eee6ff}\textbf{4.30}} & \multicolumn{1}{c}{\cellcolor[HTML]{eee6ff}\textbf{0.993}} & \multicolumn{1}{c}{\cellcolor[HTML]{eee6ff}\textbf{0.62}} &                      \\ \cline{1-7}
\multicolumn{1}{l}{}              & \multicolumn{1}{l}{}                               & \multicolumn{1}{l}{}                              & \multicolumn{1}{l}{}                              & \multicolumn{1}{l}{}                              & \multicolumn{1}{l}{}                               & \multicolumn{1}{l}{}                              &                     
\end{tabular}
\end{table}

\textbf{Vocoder Resynthesis}: To compare the waveform reconstruction capabilities of VAE and Flow-VAE, we employed both models to perform resynthesis. Metrics were computed by comparing the synthesized audio with the original audio across multiple dimensions. The results, as presented in Table~\ref{tab:objective_metrics_flowvae_vae}, indicate that the Flow-VAE model demonstrates significant advantages over the VAE model across all evaluated metrics.

\begin{table}[htbp]
\caption{\textbf{Objective indicators of TTS synthesis by VAE and Flow-VAE.}}
\label{tab:objective_metrics_tts_flowvae_vae}
\centering
\begin{tabular}{lccccc}
\toprule
                                                       &                                                    & \multicolumn{2}{c}{test-zh}                                                                         & \multicolumn{2}{c}{test-en}                                                                         \\ \cline{3-6} 
                                                       & \multirow{-2}{*}{Model}                            & \multicolumn{1}{c}{WER $\downarrow$ }                                    & SIM $\uparrow$                                     & \multicolumn{1}{c}{WER $\downarrow$}                                    & SIM $\uparrow$                                    \\ \hline
\multicolumn{1}{r}{}                                 & AR Transformer+FM+VAE                              & \multicolumn{1}{c}{0.753}                                  & 0.747                                  & \multicolumn{1}{c}{1.717}                                  & 0.633                                  \\
\multicolumn{1}{r}{\multirow{-2}{*}{zero-shot}} & AR Transformer+FM+Flow-VAE & \multicolumn{1}{c}{\textbf{0.748}} & \textbf{0.751} & \multicolumn{1}{c}{\textbf{1.639}} & \textbf{0.639} \\ \hline
\multicolumn{1}{c}{}                                 & AR Transformer+FM+VAE                              & \multicolumn{1}{c}{\textbf{0.873}} & 0.776                                  & \multicolumn{1}{c}{2.242}                                  & 0.707                                  \\ 
\multicolumn{1}{c}{\multirow{-2}{*}{one-shot}}         & AR Transformer+FM+Flow-VAE                         & \multicolumn{1}{c}{0.901}                                  & \textbf{0.782} & \multicolumn{1}{c}{\textbf{2.231}} & \textbf{0.709} \\ \bottomrule
\end{tabular}
\end{table}

\textbf{TTS Synthesis}: To assess the performance of latent features derived from VAE and Flow-VAE within a TTS framework, we trained flow matching models based on VAE latents and Flow-VAE latents, respectively on a substantial subset of our data. Following the WER and SIM evaluation methodologies from Seed-TTS ~\cite{anastassiou2024seed}, we generated test data in two inference settings: zero-shot and one-shot. The calculated WER and SIM scores are presented in Table~\ref{tab:objective_metrics_tts_flowvae_vae}.

It is worth noting that compared to the VAE model, Flow-VAE not only has advantages in the WER and SIM indicators. Upon listening to the synthesized audio, we found that Flow-VAE demonstrated significant advantages in overall stability. We encourage readers to experience through the \href{{https://minimax-ai.github.io/tts_tech_report}}{demo} link.

\section{Extensions}
\label{sec:extensions}
The disentangled and robust speaker representations learned by the integrated speaker encoder endow MiniMax-Speech with notable flexibility, facilitating its straightforward extension to various downstream applications. Because the speaker encoder captures pure vocal identity from reference audio without transcription, it provides a stable and versatile foundation upon which these extensions can be built. In this section, we detail three such extensions: (i) the control of emotional expression in synthesized speech utilizing the Low-Rank Adaptation (LoRA) technique; (ii) the generation of arbitrary and diverse vocal timbres from natural language descriptions; and (iii) professional voice cloning (PVC), a parameter efficient fine-tuning approach designed to enhance synthesis quality and fidelity for specific speakers by optimizing their associated embeddings.

\subsection{Emotion Control}

Emotional expression, conveyed through prosodic features like pitch and duration, is crucial for natural synthesized speech and is primarily modeled by the autoregressive Transformer in Minimax-Speech. We introduce a novel approach using LoRA~\cite{hu2022lora} for precise emotional control. We define discrete emotion categories, train independent LoRA modules for each using high-quality emotion-specific datasets, and dynamically load the appropriate module during inference based on user selection. This method offers higher precision and stability in emotional expression compared to natural language control.

The efficacy of this approach depends a lot on the training data, which are formatted as <reference audio, text, target emotive audio>. Reference audio provides speaker identity and establishes an emotional contrast with the target emotive audio, which the LoRA module learns to bridge. We investigated different reference audio types:

\begin{itemize}
    \item \textbf{Emotionally Congruent Reference}: Led to output emotion being overly reliant on the reference's emotion, limiting direct control.

    \item \textbf{Neutral or Random Emotion Reference}: Enabled effective control via the specified emotion category. Neutral references yielded higher expressiveness due to clearer emotional contrast, while random emotion references produced robustly natural speech with stable speaker similarity, likely by enhancing the model's ability to disentangle speaker identity from varied expressions.
\end{itemize}

To decouple synthesized emotion from lexical content, we collected multiple emotive audio samples for the same text, each with a different emotion. This trains the model to articulate identical content with varied emotional inflections, ensuring that the learned emotional rendering is independent of the text's semantic meaning.

A key advantage of this LoRA-based approach is that emotion-specific modules are trained without modifying the pre-trained Minimax-Speech core architecture. This simplifies training and deployment, preserves the original voice cloning performance, and offers excellent scalability. Experimental results show that our method achieves notable enhancements in the accuracy and naturalness of emotional expression compared to existing methodologies, producing more vivid and engaging utterances.

\subsection{Text to Voice}

Generating speech in a desired timbre with most existing TTS methods necessitates providing a reference audio sample of that specific timbre, a requirement that can limit their operational flexibility. In contrast, we introduce a T2V framework that uniquely integrates open-ended natural language descriptions with structured tag information. As a complement to the reference-audio-driven speaker encoder (which excels at cloning existing voices), this approach facilitates highly flexible and controllable timbre generation, thereby significantly enhancing the versatility of TTS systems.

Firstly, we curated a high-quality speech dataset with attributes including speech rate, gender, language, pitch, and volume. Inspired by Spark-TTS~\cite{wang2025spark}, these attributes were discretized. (e.g., pitch was divided into six bins [0, 1, 2, 3, 4, 5] according to its value in Hertz, with 0 denoting 'unknown'). These structured attributes are then combined with textual descriptions and speech data to form an aligned corpus of text-speech pairs.

Subsequently, timbre representations were extracted from AR transformer and the flow matching model. Principal Component Analysis (PCA)~\cite{mackiewicz1993principal} was employed to compress these high-dimensional features into 128 dimensions, retaining core timbre characteristics while reducing the complexity of predicting these representations. These compressed timbre representations, in conjunction with structured attributes and textual descriptions, are subsequently inputted to a compact timbre generation model. This model is trained to map natural language timbre descriptions and discrete speech attributes onto the aforementioned compressed timbre representation space. During the training phase, a random masking augmentation mechanism was introduced: key semantic words within the textual descriptions are randomly masked with a predefined probability, thereby enhancing the model's robustness with incomplete input.

This proposed framework, by combining open-ended textual descriptions with structured tag parameters, establishes a versatile timbre generation system. This system effectively unifies textual descriptions with audio-derived timbre representations for controlling timbre, empowering users to generate desired vocal characteristics using natural language (e.g., "a warm, middle-aged female voice with a slightly fast speech rate"), thereby significantly enhancing the flexibility of audio replication scenarios.

\subsection{Professional Voice Clone}

The learnable speaker encoder in the MiniMax-Speech model not only affords the model enhanced flexibility in zero-shot voice cloning tasks (due to its text-independent operation and ability to capture pure vocal identity), but also critically provides a streamlined pathway for efficient and rapid parameter fine-tuning tailored to individual speakers. Drawing inspiration from contemporary Parameter-Efficient Fine-Tuning (PEFT) methodologies ~\cite{li2021prefix,liu2021p}, we introduce a novel fine-tuning strategy. This strategy conceptualizes the conditional embedding, which encapsulates a specific speaker's vocal identity (initially derived from the speaker encoder's understanding of voice characteristics) as a set of learnable parameters. During the fine-tuning phase for a target speaker, this dedicated embedding is optimized, substituting the pre-existing speaker encoder.

Specifically, to optimize the voice timbre for any given target speaker, an initial collection of their speech data is acquired. Throughout the fine-tuning process, the autoregressive Transformer is employed as the underlying base model, and all its parameters are kept fixed (i.e., frozen). Optimization is performed exclusively on the conditional embedding associated with the target speaker, treating it as the only trainable parameter set for this adaptation. Subsequently, during the inference stage, this fine-tuned, speaker-specific conditional embedding is directly invoked to replace the real-time output generated by the standard speaker encoder.

The rationale behind PVC is to refine the speaker representation within the latent space established by the speaker encoder. Although the speaker encoder adeptly captures significant speaker information from reference audio for zero-shot voice cloning, the conditional embedding it generates for a specific speaker can be further optimized for enhanced accuracy if sufficient speech data from that speaker is available. Fine-tuning the compact conditional embedding is more tractable and offers greater flexibility for tailoring to an individual speaker compared to optimizing the entire, already well-generalized speaker encoder. Our experiments demonstrate that, with appropriate hyperparameter tuning, this PVC approach enables the model to synthesize speech exhibiting improved fidelity to the target speaker's unique timbre and superior overall perceptual quality, especially for speakers with strong accents or distinctive vocal characteristics.

This method also offers significant advantages in terms of scalability and efficiency. Because adaptation for each speaker requires optimizing only a singular vector embedding, the system readily facilitates fine-tuning and deployment for potentially thousands of distinct speakers. This is achieved without altering the foundational model's core architecture or necessitating the deployment of complete, individual models per speaker. Compared to Supervised Fine-Tuning (SFT) or even methods like LoRA, our proposed technique demonstrably curtails training complexity and reduces computational resource expenditure. This is achieved while concurrently ensuring enhancements in both the speaker similarity and the naturalness of the synthesized audio, highlighting its superior practicality and extensibility for real-world applications. For instance, its application in education allows for targeted fine-tuning to specific teacher voices, enabling the efficient generation of personalized audio content that enriches instructional materials and boosts learner engagement.

\section{Conclusion}

In this work, we have presented MiniMax-Speech, an autoregressive Transformer-based TTS model. Existing TTS methods, particularly for robust zero-shot voice cloning and high-fidelity synthesis, often face challenges such as a dependency on transcribed reference audio, which can limit cross-lingual capabilities and expressiveness, or they may struggle to achieve optimal audio quality and speaker similarity due to limitations in their generative components. To address these limitations, MiniMax-Speech introduces key innovations: its learnable speaker encoder and our novel Flow-VAE architecture integrated within a flow matching mechanism. Specifically, the learnable speaker encoder enables robust zero-shot voice cloning by extracting speaker timbre directly from reference audio, crucially without requiring accompanying text, offering superior performance in cross-lingual scenarios and a wider decoding space for generating richer and more natural prosodic variations. Concurrently, our Flow-VAE enhances the information representation power of the audio generation process, further improving both the overall audio quality and speaker similarity of the synthesized speech.  Through this combined approach, MiniMax-Speech capably supports synthesis in 32 languages. Furthermore, it has demonstrated SOTA performance on objective and subjective evaluations. Notably, this includes achieving top results on voice cloning metrics and securing the leading position on the public TTS Arena leaderboard. The extensibility granted by the speaker encoder has been showcased through applications like LoRA-based emotion control, text-driven timbre generation, and efficient professional voice cloning, establishing MiniMax-Speech as a powerful and versatile solution for high-fidelity, expressive, and controllable speech synthesis. Future work will explore further enhancements to controllability and efficiency.


\newpage 

\appendix 

\newpage

\section{Contributors} 
The contributors to the report are listed in alphabetical order as follows:

Bowen Zhang,
Congchao Guo,
Geng Yang,
Hang Yu,
Haozhe Zhang,
Heidi Lei,
Jialong Mai,
Junjie Yan,
Kaiyue Yang,
Mingqi Yang,
Peikai Huang,
Ruiyang Jin,
Sitan Jiang,
Weihua Cheng,
Yawei Li,
Yichen Xiao,
Yiying Zhou,
Yongmao Zhang,
Yuan Lu,
Yucen He

\end{document}